\documentclass{nature_modified}

\usepackage{graphicx}
\usepackage{color}
\usepackage{amsfonts,amsmath,amsthm}
\usepackage{bm}
\usepackage[colorlinks=false, pdfborder={0 0 0}]{hyperref}
\usepackage{chapterbib}

\title{Stabilizing entanglement autonomously between two superconducting qubits}

\author{S. Shankar$^{1}$, M. Hatridge$^1$, Z. Leghtas$^1$, K. M. Sliwa$^1$, A. Narla$^1$, U. Vool$^1$, S. M. Girvin$^1$, L. Frunzio$^1$, M. Mirrahimi$^{1,2}$ \& M. H. Devoret$^1$}

\newcommand{\bket}[1]{\left<#1\right>}

\newcommand{\ket}[1]{\left|#1\right>}
\newcommand{\bra}[1]{\left<#1\right|}
\newcommand{\tr}[1]{\text{Tr}\left(#1\right)}

\newcommand{\ba}{\text{\bf{a}}}

\newcommand{\bH}{\text{\bf{H}}}
\newcommand{\bD}{\text{\bf{D}}}
\newcommand{\bO}{\text{\bf{O}}}
\newcommand{\bI}{\text{\bf{I}}}
\newcommand{\brho}{\text{\boldsymbol{$\rho$}}}

\newcommand{\bsigma}{\text{\boldsymbol{$\sigma$}}}

\newcommand{\proj}[1]{\left|#1\right>\left<#1\right|}

\newcommand{\fAzero}{\omega_{A}^0}
\newcommand{\fBzero}{\omega_{B}^0}
\newcommand{\fAn}{\omega_{A}^n}
\newcommand{\fBn}{\omega_{B}^n}
\newcommand{\chiA}{\chi_{A}}
\newcommand{\chiB}{\chi_{B}}
\newcommand{\fcgg}{\omega_c^{gg}}
\newcommand{\fceg}{\omega_c^{eg}}
\newcommand{\fcge}{\omega_c^{ge}}
\newcommand{\fcee}{\omega_c^{ee}}

\newcounter{firstbib}

\begin{document}

\maketitle

\begin{affiliations}
 \item Department of Applied Physics and Physics, Yale University, New Haven, CT 06520, USA.
 \item INRIA Paris-Rocquencourt, Domaine de Voluceau, B. P. 105, 78153 Le Chesnay Cedex, France.
\end{affiliations}

\begin{abstract}
Quantum error-correction codes would protect an arbitrary state of a multi-qubit register against decoherence-induced errors\cite{Nielsen2004}, but their implementation is an outstanding challenge for the development of large-scale quantum computers. A first step is to stabilize a non-equilibrium state of a simple quantum system such as a qubit or a cavity mode in the presence of decoherence. Several groups have recently accomplished this goal using measurement-based feedback schemes\cite{Sayrin2011,Vijay2012,Riste2012b,Campagne-Ibarcq2013}. A next step is to prepare and stabilize a state of a composite system\cite{Krauter2011,Brakhane2012,Riste2013}. Here we demonstrate the stabilization of an entangled Bell state of a quantum register of two superconducting qubits for an arbitrary time. Our result is achieved by an autonomous feedback scheme which combines continuous drives along with a specifically engineered coupling between the two-qubit register and a dissipative reservoir. Similar autonomous  feedback techniques have recently  been used for qubit reset\cite{Geerlings2013} and the stabilization of a single qubit state\cite{Murch2012}, as well as for creating\cite{Barreiro2011} and stabilizing\cite{Krauter2011} states of multipartite quantum systems. Unlike conventional, measurement-based schemes, an autonomous approach counter-intuitively uses engineered dissipation to fight decoherence\cite{Poyatos1996,Kerckhoff2010,Kastoryano2011,Sarlette2011}, obviating the need for a complicated external feedback loop to correct errors, simplifying implementation. Instead the feedback loop is built into the Hamiltonian such that the steady state of the system in the presence of drives and dissipation is a Bell state, an essential building-block state for quantum information processing. Such autonomous schemes, broadly applicable to a variety of physical systems as demonstrated by a concurrent publication with trapped ion qubits\cite{Lin2013}, will be an essential tool for the implementation of quantum-error correction.
\end{abstract}

Here we implement a proposal\cite{Leghtas2013}, tailored to the circuit Quantum Electrodynamics (cQED) architecture\cite{Wallraff2004}, for stabilizing entanglement between two superconducting transmon qubits\cite{Schreier2008}. The qubits are dispersively coupled to an open cavity which acts as the dissipative reservoir. The cavity in our implementation is furthermore engineered to preferentially decay into a $50$~$\Omega$ transmission line that we can monitor on demand. We show, using two-qubit quantum state tomography and high-fidelity single-shot readout, that the steady-state of the system reaches the target Bell state with a fidelity of $67$~\%, well above the $50$~\% threshold that witnesses entanglement. As discussed in Ref.~17, the fidelity can be further improved by monitoring the cavity output and performing conditional tomography when the output indicates that the two qubits are in the Bell state. We implemented this protocol via post-selection and demonstrated that the fidelity increased to $\sim$~$77$~\%. 

Our cQED setup, outlined schematically in Fig.~1a, consists of two individually addressable qubits, Alice and Bob, coupled dispersively to a three-dimensional (3D) rectangular copper cavity. The setup is described by the dispersive Hamiltonian\cite{Nigg2012}
\begin{equation}
H/\hbar = \fAzero a^{\dagger}a + \fBzero b^{\dagger}b + \fcgg c^{\dagger}c -\alpha_A(a^{\dagger}a)^2/2-\alpha_B(b^{\dagger}b)^2/2- \chiA a^{\dagger}a c^{\dagger}c - \chiB b^{\dagger}b c^{\dagger}c,
\end{equation}
where $a$, $b$, $c$ ($a^{\dagger}$, $b^{\dagger}$, $c^{\dagger}$) are respectively the annihilation (creation) operators of Alice and Bob qubits and the cavity mode respectively. Here, $\omega_A^0$ and $\omega_B^0$ represent Alice and Bob qubit frequencies when there are no photons in the cavity, while $\omega_c^{gg}$ is the cavity frequency when both qubits are in the ground state (see Methods for experiment parameters). $\alpha_{A,B}$ are the respective qubit anharmonicities while $\chiA$ and $\chiB$ are dispersive couplings that are designed to be nearly equal. The cavity linewidth $\kappa$, is smaller than $\chi_{A,B}$, such that the system operates in the strong-dispersive limit of cQED\cite{Schuster2007}, with resolved photon number selective qubit transition frequencies ($\fAn$ and $\fBn$, where $n$ is the number of photons in the cavity). As described in Fig.~1b, by applying six continuous drives, four at qubit transitions and two at cavity transitions, an effective feedback loop is established that forces the two qubits into the Bell state $\ket{\phi_-} \equiv \left(\ket{ge}-\ket{eg}\right)/\sqrt{2}$ with zero photons in the cavity.

The feedback loop shown in Fig.~1c can be broken down into two parts that operate continuously and concurrently: the first is equivalent to a measurement process, and the second is equivalent to qubit rotations conditioned on the measurement outcome. The measurement process, embodied by two drives at the cavity frequencies, $\fcgg$ and $\fcee$, together with the approximately equal $\chi$'s, distinguishes the even-parity manifold, where the qubits are parallel, from the odd-parity manifold (where they are anti-parallel). The odd-parity manifold is conveniently described in the Bell basis $\left\{\ket{\phi_-},\ket{\phi_+}\right\}$ containing our target state $\ket{\phi_-}$. The drives, together with cavity dissipation, can be regarded as implementing a continuous projective measurement of the state of the two qubits that leaves the odd-parity manifold unaffected~\cite{Lalumiere2010,Tornberg2010}. The drives are resonant when the qubits are in $\ket{gg}$ or $\ket{ee}$, such that an average of $\bar n$ photons at $\fcgg$ or $\fcee$ continuously traverse the cavity every lifetime $1/\kappa$. On the other hand when the qubits have odd-parity, both drives are far off-resonance since $\chiA, \chiB \gg \kappa$, thus leaving the cavity almost empty of photons. As a result, the average number of photons in the cavity, the pointer variable ``observed'' by the environment\cite{Nielsen2004}, projects the even-parity manifold into $\ket{gg}$ and $\ket{ee}$, and distinguishes them from the odd-parity manifold which is left unperturbed.

The second part of the feedback loop utilizes the photon-number splitting (quantized light shifts) of the qubit transitions in the strong dispersive limit to implement conditional qubit rotations. Two drives are applied selectively at the zero-photon qubit frequencies $\fAzero$ and $\fBzero$, with amplitudes set to give equal Rabi frequencies $\Omega^0$ $\sim \kappa$ $\ll \chiA, \chiB$. The phases of these drives set, by definition, the x-axis of each qubit's Bloch sphere. The action of these drives is described by the effective unitary rotation operator $\left(a+a^{\dagger}\right)\otimes I^B+I^A\otimes\left(b+b^{\dagger}\right)$ on the two qubit subspace, where $I^A$, $I^B$ are the identity matrix. Therefore, they rotate the undesired Bell state $\ket{\phi_+,0}$ into the even-parity manifold, while the desired one $\ket{\phi_-,0}$ is left untouched. Thus, any population in $\ket{\phi_+,0}$ is eventually pumped out by the combined action of the zero-photon qubit drives and the cavity drives.

To re-pump population into the target Bell state $\ket{\phi_-, 0}$, two more drives with equal Rabi frequencies $\Omega^n$ are applied at $\fAzero-n \left(\chiA+\chiB\right)/2$ and $\fBzero-n \left(\chiA+\chiB\right)/2$ which are near the $n$-photon qubit frequencies $\fAn$ and $\fBn$ shown in Fig.~1a. The phase of the drive on Alice is set to be along the $x$-axis of its Bloch sphere while that on Bob is set to be anti-parallel to its $x$-axis resulting in the effective unitary rotation operator $\left(a+a^{\dagger}\right)\otimes I^B-I^A\otimes\left(b+b^{\dagger}\right)$ (see Methods for details on the phase control of qubit drives). Thus, as long as $n \approx \bar n$, these two drives shuffle population from $\ket{gg,n}$ and $\ket{ee,n}$ into the state $\ket{\phi_-,n}$. The latter state is unaffected by the cavity drives and therefore decays irreversibly at a rate $\kappa$ to the desired target Bell state $\ket{\phi_-,0}$.

Thus, when all six drives are turned on, the continuously operating feedback loop forces the two qubits into $\ket{\phi_-}$ even in the presence of error-inducing $T_1$ and $T_2$ processes. As shown in Extended Data Fig.~4, simulations indicate that the drive amplitudes should be optimally set to give Rabi frequencies $\Omega^0 \approx \Omega^n \approx \kappa/2$ and $\bar n \approx 3 - 4$. With these parameters, the two qubits are expected to stabilize into the target at a rate of order $\kappa/10$, which can be understood from the successive combination of transition rates, each of which is of order $\kappa$. With the experimental parameters in our implementation, the feedback  loop is expected to correct errors and stabilize the two qubits to  $\ket{\phi_-}$ with a time constant of about $1$~$\mu$s.

The fidelity of the stabilized state to the target $\ket{\phi_-}$ is determined by the competition between the correction rate $\kappa/10$ and the decoherence rates $\Gamma_1^{A,B}=1/T_{1}^{A,B}$ and $\Gamma_{\phi}^{A,B}=1/T_{\phi}^{A,B}$ that take the Alice-Bob system out of the target. Thus, to achieve a high-fidelity entangled state, the system-reservoir parameters $\chiA$, $\chiB$ and $\kappa$ have to be engineered simultaneously while maintaining  $\kappa \gg \Gamma=\textrm{max}\left(\Gamma_{1}^{A,B},\Gamma_{\phi}^{A,B}\right)$. In our present experiment, we have built on recent advances in the coherence of superconducting qubits achieved by using 3D cavities\cite{Paik2011} to obtain $\kappa/\Gamma\sim 100$, satisfying this requirement.

The protocol of the experiment consists of applying the six continuous drives for a length of time $T_S$ (see Fig.~2a) and verifying the presence of entanglement by performing two-qubit state tomography. State tomography is realized by applying one out of a set of $16$ single-qubit rotations followed by joint qubit readout\cite{Filipp2009} implemented here using single-shot measurements. As described in Fig.~1a, Extended Data Fig.~1 and Methods, the dispersive joint two-qubit readout is implemented by pulsing the cavity at $\fcgg$ and recording the cavity output for $240$~ns using a nearly quantum-limited microwave amplification chain. The first amplifier in the chain is a Josephson Parametric Converter (JPC) operated as a phase-preserving amplifier\cite{Bergeal2010a}, which performs single-shot projective readout of the state of the two qubits with a fidelity above $96$~\% in our experiment (see Extended Data Fig.~2 and Methods). The averages of the two-qubit Pauli operators are calculated by repeating the tomography $5\times10^5$ times resulting in a statistical imprecision of about $0.2$~\%.

Tomography results as a function of the duration of the stabilization interval $T_S$ are illustrated in Fig.~2b, showing the expected convergence of the system to the Bell-state $\ket{\phi_-}$. For $T_S = 0$, the Pauli operator averages $\bket{ZI} = \bket{IZ} = 0.86$ and $\bket{ZZ} = 0.72$ indicating that the system is mostly in $\ket{gg}$ when no drives are present. With increasing $T_S$, the single-qubit averages tend to zero, while the two-qubit averages $\bket{ZZ}$, $\bket{XX}$ and $\bket{YY}$ stabilize at negative values, whose sign is characteristic of $\ket{\phi_-}$. The autonomous feedback loop operation was verified with snapshots taken up to $T_S = 500$~$\mu$s, indicating that the steady-state of the two qubits remains stable for times well in excess of the time scales of decoherence processes.

The fidelity $F$ of the measured state to the target $\ket{\phi_-}$ is $F = \textrm{Tr}\left(\rho_{\mathrm{target}} \rho_{\mathrm{meas}}\right)$, where $\rho_{\mathrm{target}}=\proj{\phi_-}$ and $\rho_{\mathrm{meas}}$ is obtained from the measured set of two-qubit Pauli operators. As shown in Fig.~2c, $F$ stabilizes to $67$~\%, well above the $50$~\% threshold that indicates the presence of entanglement. This presence of entanglement is also demonstrated by the non-zero concurrence\cite{Wootters1998} $C=0.36$. The exponential rise of $F$ with a time constant of $960$~ns (see Fig.~2c inset), approximately $10$ cavity lifetimes, is in good agreement with the expected $1$~$\mu$s correction time constant of the autonomous feedback loop.

As expected, the steady-state reached by the autonomous feedback loop is impure. By analyzing the density-matrix constructed from tomography, we calculate that it contains $67$~\% weight in $\proj{\phi_-}$, while the weights in the undesired states $\proj{gg}$, $\proj{ee}$ and $\proj{\phi_+}$ are $15$~\%, $10$~\% and $8$~\% respectively. However, the reservoir (cavity output) contains information on the qubits' parity that can be exploited. Thus the state fidelity can be conditionally enhanced by passively monitoring this output and performing tomography only when the loop indicates that the qubits are in an odd-parity state (eliminating weights in even-parity states $\proj{gg}$ and $\proj{ee}$). A version of this protocol, shown in Fig.~3, is implemented in essence by passively recording the cavity output at the $\fcgg$ frequency for the last $240$~ns of the stabilization period (labeled M$_1$). A reference histogram (Fig.~3b) for M$_1$ shows Gaussian distributions separated by two standard deviations, allowing us to remove any $\proj{gg}$ present in the ensemble at the end of the stabilization period by applying an exclusionary threshold $I_m^{th}/\sigma$. We again perform single-shot tomography (M$_2$) after $100$~ns and post-select M$_2$ for outcomes M$_1= \overline{GG}$, in the process keeping $\simeq 1$~\% of counts. This conditioned tomography improves $F$ to $77$~\% ($C=0.54$) in good agreement with a simple estimate of $67/(67+10+8)=79$~\% when the weight of $\proj{gg}$ is removed. Additionally monitoring the cavity output at $\fcee$ could straightforwardly improve the conditioned fidelity to $90$~\%, achievable through modest improvements of the Josephson amplifier bandwidth. Thus, in addition to fully autonomous Bell state stabilization, we have demonstrated a proof-of-principle that real-time electronics monitoring can significantly increase the Bell-state purity.

We now address the basic imperfections of our experiment which determine the current value of the steady-state, unconditioned infidelity of $1 - 0.67 = 0.33$. First, the fidelity measured by tomography is $\simeq 0.05$ less than the theoretical steady-state value during the stabilization period, due to the $500$~ns wait time before tomography, where the state decays under the influence of $T_1$ and $T_{\phi}$. This waiting period, introduced to ensure that the single-qubit rotations during tomography are not perturbed by residual cavity photons, can be reduced with conditioned tomography ($= 100$~ns, see Fig. 3a) as there will be fewer cavity photons when the qubits are in the target state. As shown in Methods, due to this waiting period, the fidelity measured by tomography is not expected to be affected by extraneous systematic errors in calibration of single-qubit rotations (Extended Data Fig.~3). Furthermore, we also show that the fidelity is unaffected by the measurement infidelity of the readout. Instead, the sources of error are intrinsically determined by the environmental couplings inherently part of the system and its coupling to the reservoir.

One would naturally suppose that mismatch between $\chiA$ and $\chiB$ would be a dominant cause for infidelity through measurement-induced dephasing\cite{Tornberg2010}, but simulations (see Methods) indicate that the present $10$~\% mismatch contributes only $0.02$ to the infidelity. This robust property of the feedback loop is achieved by setting the cavity drives on $\fcgg$ and $\fcee$ transitions, which mitigate the measurement-induced dephasing, compared to the more straightforward irradiation between $\fcge$ and $\fceg$ proposed in usual parity measurement\cite{Lalumiere2010,Tornberg2010,Riste2013}. The dominant mechanism for infidelity turns out to be the $T_1$ and $T_{\phi}$ processes which contribute $0.12$ and $0.08$, respectively. The severity of these processes may actually be enhanced in the presence of cavity drives which tend to shorten $T_1$ in transmon qubits\cite{Slichter2012}. The remaining infidelity can possibly be attributed to this shortening of $T_1$, in combination with the presence of $\ket{f}$ (second excited state) population in each qubit due to finite qubit temperature. Overall, the dominant sources of infidelity ($T_1$, $T_{\phi}$) are likely to be mitigated in future experiments by expected improvements in qubit coherence (see Methods), which will allow a larger ratio $\kappa/\Gamma$ and thus fidelities in excess $90$~\%.

In conclusion, we have demonstrated the stabilization of two-qubit entanglement that makes a Bell state available for indefinite time using a completely autonomous protocol. While the fidelity to the Bell state doesn't presently exceed that needed for Bell's inequality violation, straightforward improvements to the setup should allow it\cite{Leghtas2013}. In addition to sufficiently coherent qubits, the resources required, consisting of matching dispersive couplings and six pure tones sent through the same input line, are modest in comparison with the hardware that would be needed in a conventional measurement-based scheme to stabilize $\ket{\phi_-}$. While the pure tones can be generated using microwave modulation techniques from three sources only, the tolerance to imperfections in the matching of the couplings amounts to about $10$~\%, easily achieved in superconducting qubit design. A primary virtue of our protocol is that it can be extended to larger systems as it only assumes that (a) the system Hamiltonian can be precisely engineered (a general requirement for all quantum information implementations) and (b) that abundant, off-the-shelf room temperature microwave generators are available. Moreover, the protocol can take advantage of any available high-fidelity readout capability by passively monitoring the cavity output(s), enabling purification by real-time conditioning as demonstrated by our $30$~\% reduction in state infidelity using a Josephson amplifier. 

Therefore autonomous feedback is uniquely suited to take advantage of all the state-of-the-art hardware available, and is an ideal platform to construct more complicated protocols. Possible avenues for future experiments include implementing a ``compound'' Bell-state stabilization protocol, using a four-qubit quantum register operating two-pairwise autonomous stabilization stages complemented by an entanglement distillation step\cite{Reichle2006}. The quantum engineering concept implemented in our experiment could also be applied to autonomously stabilize a coherent two-state manifold of Schr\"{o}dinger cat states of a superconducting cavity\cite{Leghtas2012}, thus possibly achieving a continuous version of quantum error-correction of a qubit.

\section*{Methods summary}
Alice and Bob are single-junction 3D transmons with qubit frequencies $\fAzero/2\pi=5.238$~GHz, $\fBzero/2\pi= 6.304$~GHz, anharmonicities $\alpha_A/2\pi = 220$~MHz, $\alpha_B/2\pi = 200$~MHz, relaxation times $T_1^A = 16$~$\mu$s, $T_1^B = 9$~$\mu$s and pure dephasing times $T_{\phi}^{A} = 11$~$\mu$s, $T_{\phi}^{B} = 36$~$\mu$s. They are coupled to a rectangular cavity ($\fcgg/2\pi = 7.453$~GHz) with nearly equal dispersive couplings $\chiA/2\pi = 6.5$~MHz, $\chiB/2\pi = 5.9$~MHz, that are larger than the cavity linewidth, $\kappa/2\pi = 1.7$~MHz. The setup is mounted to the base of a dilution refrigerator (Extended Data Fig.~1) and controlled using heavily attenuated and filtered microwave lines. The room temperature microwave setup generates all microwave drives in a manner which is insensitive to drifts in generator phases. Single-shot joint readout with fidelity of $96$~\% was performed using a JPC (Extended Data Fig.~2). We checked by performing tomography of Clifford states that the measured fidelity of the Bell-state was not significantly altered by systematic errors in single-qubit rotations and measurements (Extended Data Fig.~3). The choice of drive amplitudes used for Bell-state stabilization was guided by Lindblad master equation simulations (Extended Data Fig.~4). These simulations also provided an error-budget analysis for the steady state infidelity suggesting that the dominant source of infidelity is the finite $T_1$ and $T_{\phi}$.

\begin{addendum}
 \item[Acknowledgements] Facilities use was supported by the Yale Institute for Nanoscience and Quantum Engineering (YINQE) and the National Science Foundation (NSF) MRSEC DMR 1119826. This research was supported by the Office of the Director of National Intelligence (ODNI), Intelligence Advanced Research Projects Activity (IARPA) W911NF-09-1-0369, by the U.S. Army Research Office W911NF-09-1-0514 and by the NSF DMR 1006060 and DMR 0653377. MM acknowledges partial support from the Agence National de Recherche under the project EPOQ2 ANR-09-JCJC-0070. SMG and ZL acknowledge support from the NSF DMR 1004406. All statements of fact, opinion or conclusions contained herein are those of the authors and should not be construed as representing the official views or policies of IARPA, the ODNI, or the U.S. Government.
 \item[Correspondence] Correspondence and requests for materials
should be addressed to S. Shankar~(email: shyam.shankar@yale.edu) or M. H. Devoret~(email: michel.devoret@yale.edu)
\end{addendum}

\clearpage
\begin{figure}
\centering
\includegraphics{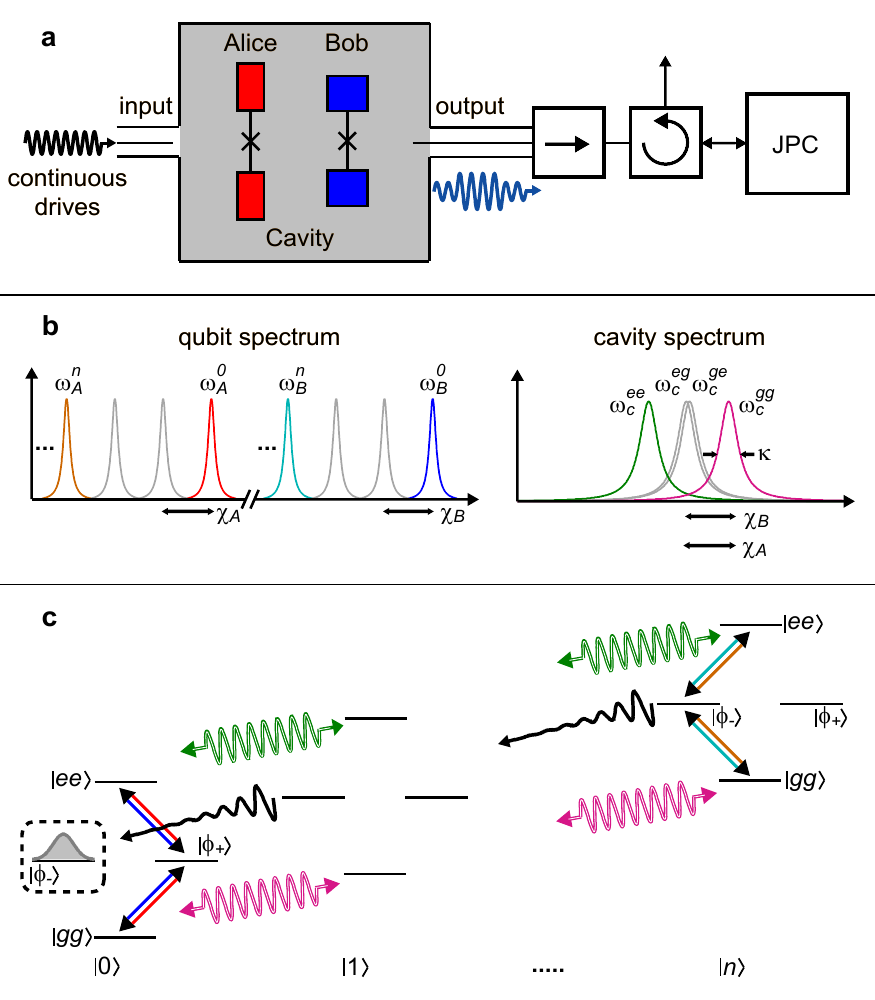}
\end{figure}
\textbf{Figure 1: Bell state stabilization : setup schematic and frequency landscape of autonomous feedback loop.}
\textbf{a.} The qubits (Alice, Bob) are coupled to the fundamental mode of a 3D cavity. Six continuous drives applied to the cavity input stabilize the Bell state $\ket{\phi_-,0}$. The cavity output is schematically shown to jump between low and high amplitude when the qubits are in the desired Bell state or not. The output is monitored by a quantum-limited amplifier (JPC).
\textbf{b.} Spectra of the qubits and cavity coupled with nearly equal dispersive shifts ($\chiA$, $\chiB$). Cavity linewidth is $\kappa$. Colors denote transitions which are driven to establish the autonomous feedback loop.
\textbf{c.} Effective states of the system involved in the feedback loop. Qubit states consist of the odd-parity states in the Bell basis $\left\{\ket{\phi_-},\ket{\phi_+}\right\}$ and the even-parity computational states $\left\{\ket{gg},\ket{ee}\right\}$. Cavity states, arrayed horizontally, are the photon number basis kets $\ket{n}$. Sinusoidal double lines represent the two cavity tones whose amplitudes create on average $\bar n$ photons in the cavity when the qubits are in even-parity states. The cavity level populations are Poisson distributed with mean $\bar n$ and we just show $\ket{n}$ such that $n\approx\bar n$. Straight double lines represent four tone on qubit transitions. Collectively, the six tones and the cavity decay (decaying sinusoidal lines) drive the system towards the ``dark'' state $\ket{\phi_-,0}$, which builds up a steady-state population.

\clearpage
\begin{figure}
\centering
\includegraphics{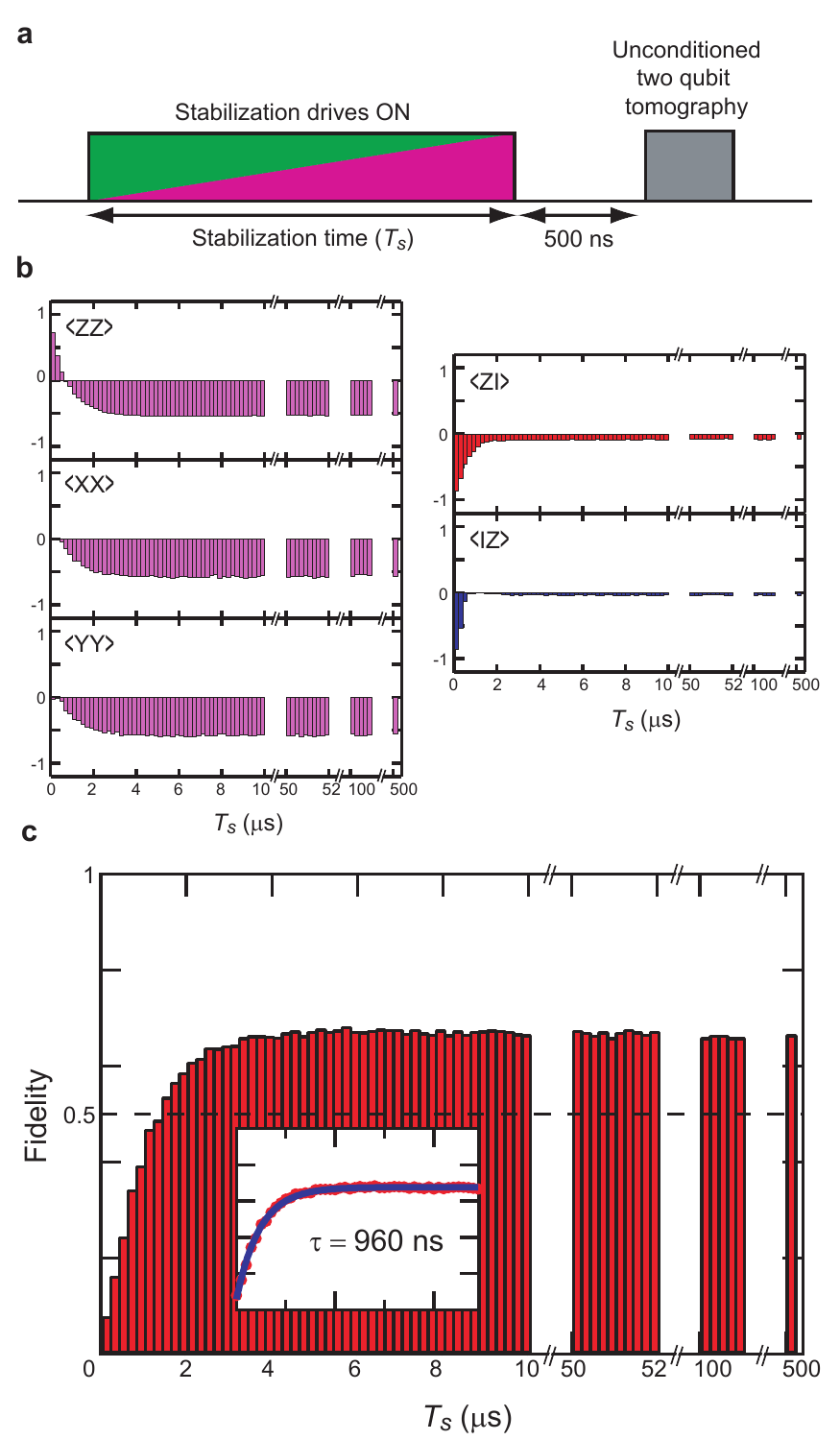}
\end{figure}
\textbf{Figure 2: Convergence of two qubit-state to the target Bell state.}
\textbf{a.} The six drives are turned on during the stabilization period for time $T_S$. Next, the drives are turned off and the system is left idle for $500$~ns, allowing any remaining cavity photons to decay away. Finally, two-qubit tomography is performed using single-qubit rotations followed by single-shot joint readout. The system is then allowed to reach thermal equilibrium by waiting at least 5T$_1$ before repetition.
\textbf{b.} Time variation of the relevant Pauli operator averages, showing the system's evolution from thermal equilibrium (nearly $\ket{gg}$) towards  $\ket{\phi_-}$. The system remains in this steady-state for arbitrarily long times, as demonstrated by data acquired at $T_S = 50$, $100$ and $500$~$\mu$s.
\textbf{c.} Fidelity ($F$) to the target state $\ket{\phi_-}$ versus stabilization time $T_S$. The dashed line at $50$~\% is the entanglement threshold. The fidelity converges to $67$~\% with a time constant of about $10$ cavity lifetimes, in good agreement with the theoretical prediction  (see inset showing $F$ from $T_S = 0$ to $10$~$\mu$s as red circles, with fit to exponential dependence as blue line).

\clearpage
\begin{figure}
\centering
\includegraphics{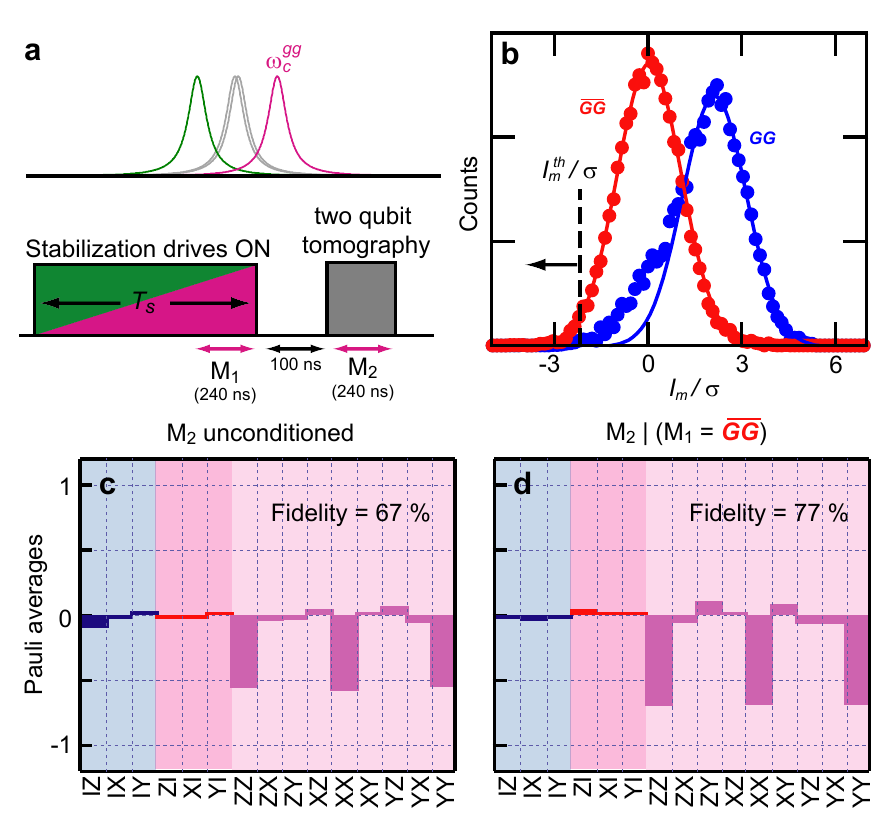}
\end{figure}
\textbf{Figure 3: Fidelity improved by monitoring the feedback loop.}
\textbf{a.} Pulse sequence consisting of a $T_S = 10$~$\mu$s period followed by two-qubit tomography. Here, the cavity output at $\fcgg$ is recorded during the last $240$~ns of the stabilization period (M$_1$). The outcomes obtained during M$_1$ are used to condition the tomography in post-processing. After waiting $100$~ns for any cavity photons to decay away, two-qubit state tomography is performed using a second $240$~ns long measurement (M$_2$) similar to that used in the unconditioned tomography (Fig.~2). 
\textbf{b.} Reference histogram for M$_1$, with qubits prepared in thermal equilibrium (denoted as $GG$) and after a $\pi$ pulse on Alice (denoted as $\overline{GG}$). The standard deviation $\sigma$ of the Gaussian distributions scales the horizontal axis of measurement outcomes $I_m$. 
\textbf{c.} Complete set of Pauli operator averages measured by tomography without conditioning, as in Fig.~2, showing a fidelity of $67$~\% to $\ket{\phi_-}$.
\textbf{d.} Tomography conditioned on M$_1$ being $\overline{GG}$, that is outcomes $I_m/\sigma$ during M$_1$ $< I_m^{th}/\sigma = -2.2$, resulting in an increased fidelity of $77$~\%.
\clearpage

\noindent{\textbf{\Large Methods}}

\section{Qubit-cavity implementation}
The two transmon qubits were fabricated with double-angle evaporated Al/AlO$_x$/Al Josephson junctions, defined using the bridge-free e-beam lithography technique\cite{Lecocq2011,Rigetti2009}, on double-side-polished $3$~mm by $10$~mm chips of c-plane sapphire. They were coupled to the TE101 mode of a rectangular copper cavity. The room-temperature junction resistances, (Alice: $7.5$~k$\Omega$ , Bob: $5.6$~k$\Omega$), antenna pad dimensions (Alice: $1.4$~mm by $0.2$~mm, Bob: $0.68$~mm by $0.36$~mm) and cavity dimensions ($35.6$~mm by $21.3$~mm by $7.6$~mm) were designed using finite-element simulations and black-box circuit quantization analysis\cite{Nigg2012} to give Alice and Bob qubit frequencies $\fAzero/2\pi=5.238$~GHz, $\fBzero/2\pi= 6.304$~GHz, qubit anharmonicities $\alpha_A/2\pi = 220$~MHz, $\alpha_B/2\pi = 200$~MHz, cavity frequency $\fcgg/2\pi = 7.453$~GHz, and nearly equal dispersive couplings $\chiA/2\pi = 6.5$~MHz, $\chiB/2\pi = 5.9$~MHz. The cavity was coupled to input and output transmission lines with quality factors $Q_{IN}\sim 100,000$ and $Q_{OUT}=4,500$ such that its linewidth, $\kappa/2\pi = 1.7$~MHz, was set predominantly by the $Q_{OUT}$.

As shown in the experiment schematic (Fig.~ED1), the cavity and the JPC setup was mounted to the base-stage of a cryogen-free dilution refrigerator (Oxford Triton200). As is common practice for superconducting qubit experiments, the cavity and JPC were shielded from stray magnetic fields by aluminum and cryogenic $\mu$-metal (Amumetal A4K) shields. The input microwave lines going to the setup were attenuated at various fridge stages and filtered using commercial $12$~GHz reflective, low-pass and home-made, lossy Eccosorb filters. The attenuators and filters serve to protect the qubit and cavity from room temperature thermal noise and block microwave or optical frequency signals from reaching the qubit. The output line of the fridge consisted of reflective and Eccosorb filters as well as two cryogenic isolators (Quinstar CWJ1019K) at base for attenuating noise coming down from higher temperature stages. In addition, a cryogenic HEMT amplifier (Low Noise Factory LNF-LNC7\_10A) at the $3$~K stage provided $40$~dB of gain to overcome the noise added by the following room-temperature amplification stages.

Relaxation times were measured to be $T_1^A = 16$~$\mu$s, $T_1^B = 9$~$\mu$s and coherence times measured with a Ramsey protocol were $T_{2}^{A}= 8$~$\mu$s, $T_{2}^{B}=12$~$\mu$s, resulting in dephasing times $T_{\phi}^{A} = 11$~$\mu$s, $T_{\phi}^{B} = 36$~$\mu$s. Black-box quantization analysis of the qubit-cavity system suggest that the relaxation times were limited by the Purcell effect\cite{Houck2008}. Coherence times did not improve using an echo pulse suggesting that they were limited by thermal photons present in the fundamental and higher modes of the cavity\cite{Sears2012} as well as non-zero qubit temperature ($\sim 75$~mK).

\section{Control of stabilization drives}
The room-temperature setup must generate and control microwave tones in a manner such that the experiment is insensitive to drifts in the phase between microwave sources over the time-scale of the experiment. While all sources are locked to a common rubidium frequency standard (SRS FS725), they drift apart in phase on a time-scale of a few minutes. Therefore, for example, the four Rabi drives on the qubits during Bell state stabilization cannot be produced by four separate sources as the phase of the drives need to be controlled precisely. The phase drift was eliminated by using one microwave source per qubit and generating the desired frequencies using single-sideband modulation (see Extended Data Fig.~1). The qubit drives were produced by sources $\textsf{f}_{\textsf{Alice}}$ and $\textsf{f}_{\textsf{Bob}}$ (Vaunix Labbrick LMS-802), set $100$~MHz below the respective zero-photon qubit frequencies. For Bell-state stabilization, these tones were mixed by IQ mixers (Marki IQ4509) with $100$~MHz and $82$~MHz sinewaves produced from a Tektronix AWG5014C arbitrary waveform generator (AWG). The mixer outputs at the desired frequencies can be expressed mathematically as $A_A^0\textrm{cos}\left(\omega_A^0 t+\phi_A^{\textrm{arb}}+\phi_A^0\right)$, $A_B^0\textrm{cos}\left(\omega_B^0 t+\phi_B^{\textrm{arb}}+\phi_B^0\right)$, $A_A^n\textrm{cos}\left(\omega_A^n t+\phi_A^{\textrm{arb}}+\phi_A^n\right)$ and $A_B^n\textrm{cos}\left(\omega_B^n t+\phi_B^{\textrm{arb}}+\phi_B^n\right)$. Here $\phi_A^{\textrm{arb}}$, $\phi_B^{\textrm{arb}}$ are the arbitrary phases of the microwave sources $\textsf{f}_{\textsf{Alice}}$ and $\textsf{f}_{\textsf{Bob}}$ which can drift during an experiment, while $\phi_A^0$, $\phi_B^0$, $\phi_A^n$ and $\phi_B^n$ are set by the AWG as well as the length of cables going to the qubit/cavity system, and are therefore fixed over the course of the experiment. The relationship between the drive phases required for the stabilization protocol is $\phi_A^0-\phi_A^n=\phi_B^0-\phi_B^n+\pi$. This is achieved in experiment by fixing $\phi_A^0$, $\phi_A^n$, $\phi_B^0$ and sweeping $\phi_B^n$.

The cavity drives for stabilization were generated by two sources  (Agilent E8267 and N5183), $\textsf{f}_\textsf{c}^{\textsf{GG}}$ set to $\fcgg$ and $\textsf{f}_\textsf{c}^{\textsf{EE}}$ set to $\fcee=\left(\fcgg-\chiA-\chiB\right)$. These drives could potentially also be produced using a single microwave source and single-sideband modulation, however this was not done as control over the phase of these drives was not important for the stabilization protocol.

\section{Joint-readout implementation with JPC}
The joint-readout of the qubits used for tomography, was implemented with high-fidelity single-shot measurements\cite{Hatridge2013}, by pulsing the cavity input for $500$~ns using the source $\textsf{f}_\textsf{c}^{\textsf{GG}}(\textsf{msmt})$ (Agilent N5183) set at $\fcgg$ (see Fig.~ED1). The transmitted microwave pulse was directed via two circulators (Quinstar CTH1409) to the JPC amplifier, reflected with gain, then amplified at $3$~K followed by further signal processing at room temperature. The JPC was biased at $\fcgg$ to provide a reflected power gain of $20$~dB in a bandwidth of $6$~MHz. A noise-rise of $6$~dB was recorded when the amplifier was on versus off implying that $80$~\% of the noise measured at room-temperature was amplified quantum fluctuations originating from the base-stage of the fridge.

The circulators also provide reverse-isolation which prevents amplified quantum fluctuations output by the JPC from impinging on the cavity and causing dephasing. In our experiment, the $T_2$ of the qubits was found to reduce to $3$~$\mu$s when the amplifier was turned on, suggesting that either this reverse-isolation was insufficient or that the pump tone was accidentally aligned with a higher mode of the cavity. Therefore, the amplifier was turned on $100$~ns before the cavity pulse was applied and turned off $1$~$\mu$s after the cavity rung down. This pulsing of the JPC ensured that the excess dephasing was absent during the stabilization period of the experiment described in the main text. Rather, it was present only during the tomography phase, when it was less important.

The output of the fridge at $\fcgg$ had to be shifted to radio-frequencies ($< 500$~MHz) before it could be digitized using commercial hardware. This processing was performed in a manner that was insensitive to drifts in digitizer offsets and generator phases over the timescale of the experiment. As shown in Fig.~ED1, the fridge output was demodulated using an image-reject mixer (Marki IRW0618) with a local oscillator ($\textsf{f}_\textsf{c}^{\textsf{GG}}(\textsf{msmt})+50$~MHz, Agilent N5183), set $50$~MHz above $\textsf{f}_\textsf{c}^{\textsf{GG}}(\textsf{msmt})$, to produce a signal centered in frequency domain at $50$~MHz. A copy of the cavity input that did not pass through the dilution refrigerator was also demodulated to give a reference copy for comparison. The signal and reference were digitized (in an analog-to-digital converter Alazar ATS9870) and finally digitally demodulated in the PC to give in-phase and quadrature signals $\left(I(t),Q(t)\right)$. This room-temperature analog and digital signal processing ensured that the measured $\left(I(t),Q(t)\right)$ were insensitive to drifts in digitizer offsets and generator phases.

A $240$~ns section of the cavity response after ring-up was averaged to give a measurement outcome $\left(I_m,Q_m\right)$. Reference histograms along the $I_m$ axis are shown in Fig.~ED2a. These reference histograms were produced using qubits initialized in $\ket{gg}$, by a heralding measurement\cite{Johnson2012}, with a fidelity $99.5$~\%. The histogram labeled $GG$ was recorded using this initialized state, while that labeled $\overline{GG}$ was recorded after a $\pi$ pulse on Alice. Similar $\overline{GG}$ histograms could be produced after a $\pi$ pulse on Bob or on both qubits. The two histograms are fitted with Gaussian distributions; the $GG$ histogram has a standard deviation that is smaller by approximately a factor of $0.75$ smaller, due to amplifier saturation, resulting in a larger amplitude. However, the area under each histogram is identical, as expected. The different standard deviations imply that the threshold $\left(I_m^{th}/\sigma\right)$ distinguishing outcomes associated with $GG$ from those associated with $\overline{GG}$ can not be set symmetrically between the two distributions.  Rather, $I_m^{th}/\sigma=5$,  shifted towards $GG$, ensuring that the error induced by the overlap of the distributions are equal. Since the Gaussians are separated by $5.5$ standard deviations, the indicated threshold would imply a readout fidelity of $99.5$~\% calculated from the overlap of the distributions.

These well-separated Gaussian distributions indicate that the readout implements a close to ideal measurement of the observable $\proj{gg}$. However, excess counts are observed in $\overline{GG}$ when the qubits are prepared in $\ket{gg}$ and vice-versa, due to $T_1$ events as well as transitions induced by the measurement tone. These errors reduce the fidelity from that calculated simply from the overlap of the two distributions. Thus, the total measurement fidelity of the observable $\proj{gg}$, summarized by the diagram in Fig.~ED2b, is found to be $96$~\% for the state $\ket{gg}$ and $97$~\% for states $\ket{ge}$, $\ket{eg}$ and $\ket{ee}$.

\section{Calibration of systematic errors in tomography}
Standard two-qubit tomography is performed by applying one out of a set of four single-qubit rotations ($\textrm{Id}$, $\textrm{R}_x(\pi)$, $\textrm{R}_x(\pi/2)$, $\textrm{R}_y(\pi/2)$) on each qubit followed by the readout of the observable $\proj{gg}$\cite{Filipp2009,Chow2010}. To implement these rotations, the AWG shapes a $100$~MHz sinewave with a $6\sigma$ long Gaussian envelope having $\sigma=12$~ns and DRAG correction\cite{Motzoi2009,Reed2013}. The output of the IQ mixers (Fig.~ED1) are thus Gaussian pulses resonant on the zero-photon qubit transition frequencies. The resulting set of $16$  measured observables are expressed in the Pauli basis and the Pauli operator averages are calculated by matrix inversion. The density matrix constructed from the Pauli operator averages is then used to calculate the fidelity of the measured state to a desired target, as described in the main text.

The ability of the tomography to faithfully represent the state of the two qubits is limited by systematic errors in the readout, in single-qubit rotations and due to qubit decoherence. While these errors have been individually reduced below a percent in superconducting qubit experiments\cite{Devoret2013}, they are in the few percent range in our system due to insufficient combined optimization. We have estimated a worst-case combined effect of these errors on the tomography by preparing the qubits in one of $36$ possible two-qubit Clifford states and then performing tomography to extract the fidelity to the target state. The fidelity of the measured state to the target, shown in Fig.~ED3, varies from a maximum of $94$~\% for the state $\ket{-Z,-Z}$ which is least susceptible to errors from relaxation and decoherence, to a minimum of $87$~\% for the state $\ket{+Y,+Z}$ which is among the most susceptible. The average fidelity of $90$~\% across all $36$ states is in good agreement with that expected from the aggregate of few percent errors arising from readout, rotations and decoherence. This substantially higher fidelity than the maximum fidelity to the Bell-state we estimate in the main text, leads us to believe that systematic errors in the tomography do not significantly alter the maximum measured fidelity of $77$~\% to $\ket{\phi_-}$.

\section{Simulation of the stabilization protocol}
The stabilization protocol consists of 6 drives whose amplitudes need to be optimized for maximum fidelity.  We now describe simulations which suggest that this is not such a daunting task as the continuous-drive protocol is robust against modest errors in the amplitudes. As described in the theory proposal\cite{Leghtas2013}, the dynamics of the system comprising two qubits, coupled to the cavity (the reservoir) in the presence of drives, qubit decay and qubit dephasing can be simulated using the Lindblad master equation
\begin{equation*}
 	\frac{d}{dt}\brho(t)=-\frac{i}{\hbar}[\bH(t),\brho(t)]+\kappa\bD[\ba]\brho(t)+\sum_{j=A,B}\left({\frac{1}{
 	T_1^{j}}\bD[\bsigma_-^{j}]\brho(t)+\frac{1}{2 T_{\phi}^{j}}\bD[\bsigma_z^{j}]\brho(t)}\right)\notag,
 \end{equation*}
where
 \begin{eqnarray*}
 \bH(t)&=&\left(\chi_A\frac{\bsigma_z^A}{2}+\chi_B\frac{\bsigma_z^B}{2}\right) \ba^\dag\ba
 +2\epsilon_c\cos\left(\frac{\chi_A+\chi_B}{2} t\right)\left(\ba+\ba^\dag\right)+\Omega^0\left(\bsigma_x^A+\bsigma_x^B\right)\\
 &&+\Omega^{n}\left(e^{-i n\frac{\chi_A+\chi_B}{2}t}\left(\bsigma_+^A-\bsigma_+^B\right)+\text{c.c.}\right),
 \end{eqnarray*}
 is the Hamiltonian of the driven system in the rotating frame of the two qubits ($\fAzero$, $\fBzero$) and the cavity mode $\left(\left\{\omega_c^{gg}+\omega_c^{ee}\right\}/2\right)$. The qubits are considered to be perfect two-level systems rather than anharmonic oscillators assumed in the main text, and therefore we use Pauli operators $\bsigma_z$, $\bsigma_x$, $\bsigma_+=\bsigma_x+i\bsigma_y$. The qubit dephasing rate is $1/T_\phi^{A,B}=1/T_2^{A,B}-1/2T_1^{A,B}$, while the Lindblad super-operator is defined for any operator $\bO$ as $\bD[\bO]\brho=\bO\brho\bO^\dag-\frac{1}{2}\bO^\dag\bO\brho-\frac{1}{2}\brho\bO^\dag\bO$. $\epsilon_c$ is the amplitude of the drive on the cavity which is taken equal to $\kappa\sqrt{\bar n}/2$ where $\bar n$ is the number of photons circulating in the cavity. The Lindblad equation is solved numerically for $\brho(t)$ assuming $\brho(0)=\proj{gg}$. The steady-state fidelity to $\ket{\phi_-}$ is estimated as $\tr{(\ket{\phi_-}\bra{\phi_-}\otimes \bI_c) \brho(\infty)}$. The system was empirically found to have reached steady-state at $t = 10~\mu\textrm{s}$, so $\brho(\infty)$ is taken to be $\brho(10~\mu\textrm{s})$.

The drive amplitudes are swept in the simulation to optimize fidelity; a representative result shown in Fig.~ED4 for our system characteristics, indicates that a broad range of cavity drive amplitudes above $3$ photons and Rabi drive amplitudes above $\kappa/2$ should lead to fidelities around $70$~\%. The dependence of the fidelity on drive amplitudes can be qualitatively understood as follows. As discussed in the main text, the two cavity drives perform a quasi-parity measurement of the state of the qubits. The parity measurement rate is $\bar n \kappa/2$ for $\chi_A, \chi_B \gg \kappa$, which increases with cavity drive amplitude. Thus the fidelity is smaller at low $\bar n$ due to the slow measurement of parity compared with the error rate induced by decoherence. On the other hand, the fidelity drops at high $\bar n$ due to the unwanted dephasing between $\ket{\phi_-}$ and $\ket{\phi_+}$ induced by the mismatch between $\chi_A$ and $\chi_B$. Next, we see that the Rabi rates required for highest fidelity increase with $\bar n$. This effect arises from a quantum Zeno-like competition\cite{Itano1990} between the parity measurement which pins the qubits in the odd or even parity subspaces and the Rabi drives that try to induce transitions between these subspaces. The ratio of the rates of these processes is the quantum Zeno parameter, which must not be too large in order to ensure that the photon number selective Rabi drives correct the system fast enough. For our optimal parameters of $\bar n = 3$ and $\Omega^0 = \Omega^n = \kappa/2$, the quantum Zeno parameter is 3, not much greater than 1. In this intermediate regime, the feedback loop does not respond to errors through fully resolved discrete quantum jumps between the various states, but rather through a quasi-continuous evolution. Moreover as seen in Fig.~ED4, this continuous feedback strategy is insensitive to small errors in setting the drive amplitudes, a favorable quality for the experimental realization.

In the experiment, the drive amplitudes for the cavity and zero-photon qubit transitions were pre-calibrated with Ramsey and Rabi experiments so that they could be set to $\bar n = 3$, $\Omega^0 = \kappa/2$. On the other hand, $\Omega^n$ and the phase of the $n$-photon Rabi drives can not be easily calibrated; instead they were individually swept till the fidelity was maximized. As a final optimization, $\bar n$ and $\Omega^0$ were also swept; the fidelity improved by $1$--$2$~\% for $\bar n = 3.7$ and $\Omega^0 = \kappa/2$, marginally different from the originally chosen parameters. Overall, we found that the drive amplitudes could be varied by about $20$~\% without reducing the fidelity by more than $1$~\%. Thus, this result as well as the good agreement for the steady-state maximum fidelity of $67$~\% indicates that the Lindblad simulation captures most of the physics of our experiment.

\section{Sources of steady-state infidelity}
The Lindblad simulation provides an error budget analysis for the steady-state infidelity, indicating directions for improvement. We first set $\chiA=\chiB=5.9$~MHz and $T_1=T_2=\infty$, and then individually introduce the imperfections into the simulation. The ideal fidelity with drive amplitudes $\bar n = 3$, $\Omega^0=\Omega^n=\kappa/2$ is $97$~\%, limited by the finiteness of $\bar n$. Introducing the $\sim 10$~\% $\chi$ mis-match, reduces this fidelity by only $2$~\%, indicating the robustness of the protocol to the difference between $\chiA$ and $\chiB$. On the other hand, individually adding $T_1$ and $T_{\phi}$ processes reduces the fidelity by $12$~\% and $8$~\% respectively. Thus, we find that the dominant sources of infidelity are the decoherence processes inherent to the qubits and their coupling to the environment.

The $T_1$ of the qubits are believed to be Purcell-limited\cite{Houck2008} (implying $\kappa T_1 = \textrm{constant}$) and potentially could be a factor of $10$ longer in the 3D cQED architecture\cite{Paik2011}. However, this improvement cannot be achieved by reducing $\kappa$, as that would concurrently reduce the feedback correction time, and thus the steady-state fidelity. Instead, $T_1$ must be improved using a Purcell filter\cite{Reed2010}, which results in a larger $\kappa T_1$ and thus an overall improvement to the fidelity. Such a filter has been implemented in the 3D architecture in our group recently and will be an immediate upgrade to the current setup.

The current limit on $T_{\phi}$ is believed to be set by dephasing arising from thermal photons in the fundamental and higher modes of the cavity. This dephasing is given by $\kappa T_{\phi} \simeq 1/n_{\mathrm{th}}$, where $n_{\mathrm{th}}$ is the thermal occupancy of the cavity modes. The calculated $n_{\mathrm{th}} = 10^{-2}$ to give our $T_2$, can be reduced by at-least an order of magnitude\cite{Sears2012}, allowing larger $\kappa T_{\phi}$ and thus an improved fidelity.

Other sources of infidelity are $\sim 4$~\% $\ket{f}$ (second excited states) population of the qubits due to finite temperature, as well as undesired qubit transitions induced by the cavity drives\cite{Slichter2012} that shorten $T_1$. The $\ket{f}$ state population could be reduced in future experiments by additional drives on the $\ket{e} \leftrightarrow \ket{f}$ transitions. On the other hand, the $T_1$ shortening remains an insufficiently understood effect which requires further investigation. Nevertheless, these effects are not likely to limit the fidelity by more than $10$~\%.

\clearpage
\begin{figure}
\centering
\includegraphics[scale=0.9]{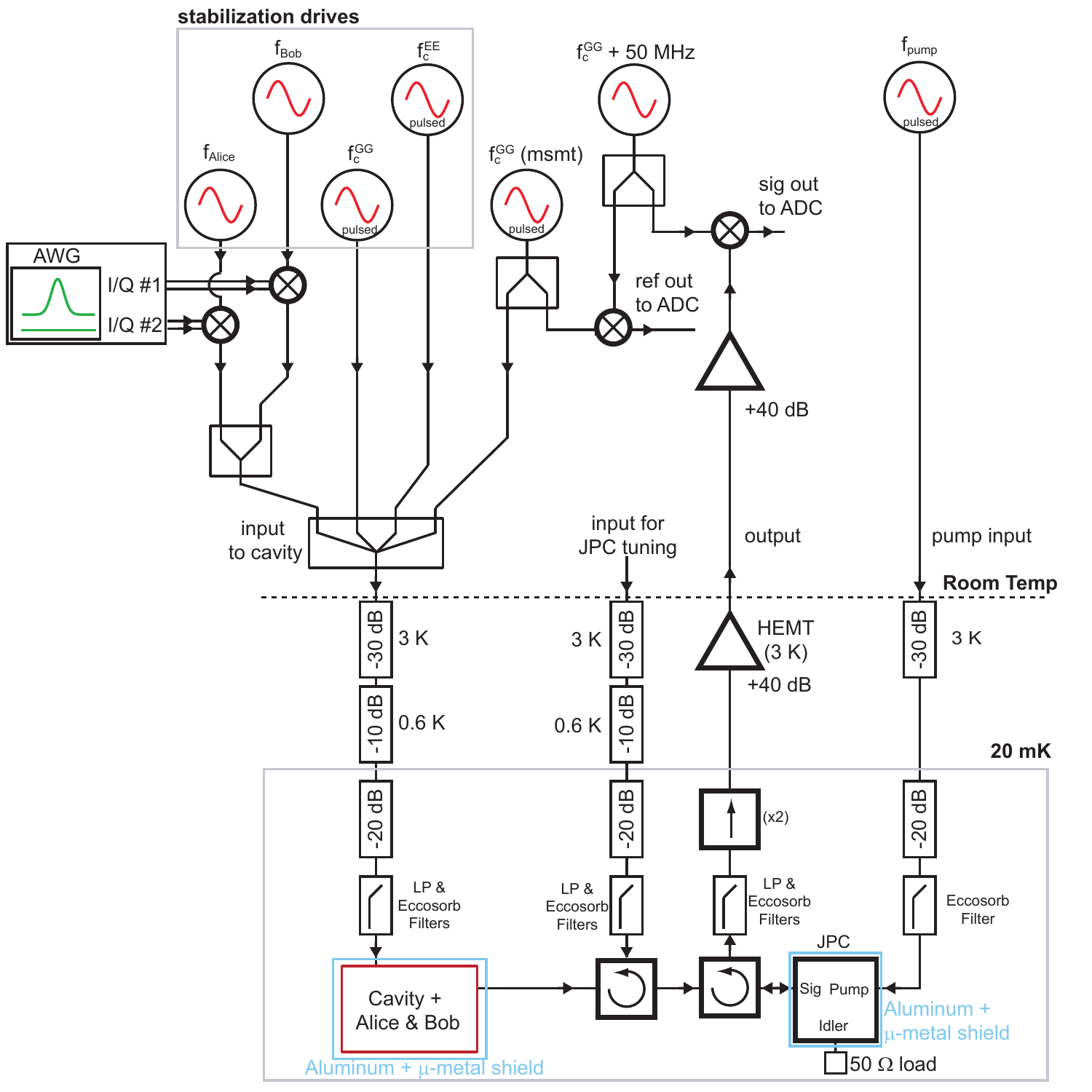}
\end{figure}
\textbf{Extended Data Figure 1: Experiment schematic}  The qubit-cavity setup as well as the JPC amplifier is mounted on the base stage of a dilution refrigerator (bottom of diagram) which is operated below $20$~mK. The room-temperature setup consists of electronics used for qubit control (top left) and for qubit measurement (top right). The experiment is controlled by an arbitrary waveform generator (AWG) which produces analog waveforms and also supplies digital markers (not shown) to the pulsed microwave sources. The drives for stabilization and qubit control are generated from four microwave sources in the present experiment though the two cavity drives  $\textsf{f}_\textsf{c}^{\textsf{GG}}$ and $\textsf{f}_\textsf{c}^{\textsf{EE}}$ could be produced in principle from the same source. These drives were combined with a measurement drive and sent through filtered and attenuated lines to the cavity input at the base of the fridge.	 The cavity output is directed to the signal port of a JPC, whose idler is terminated in a $50$~$\Omega$ load. The JPC is powered by a drive applied to its pump port. The fridge input labeled ``for JPC tuning'' is used solely for initial tune up and is terminated during the stabilization experiment. The cavity output signal is amplified in reflection by the JPC and then output from the fridge after further amplification. The output signal is demodulated at room temperature and then digitized by an analog-to-digital converter along with a reference copy of the measurement drive. 

\clearpage
\begin{figure}
\centering
\includegraphics{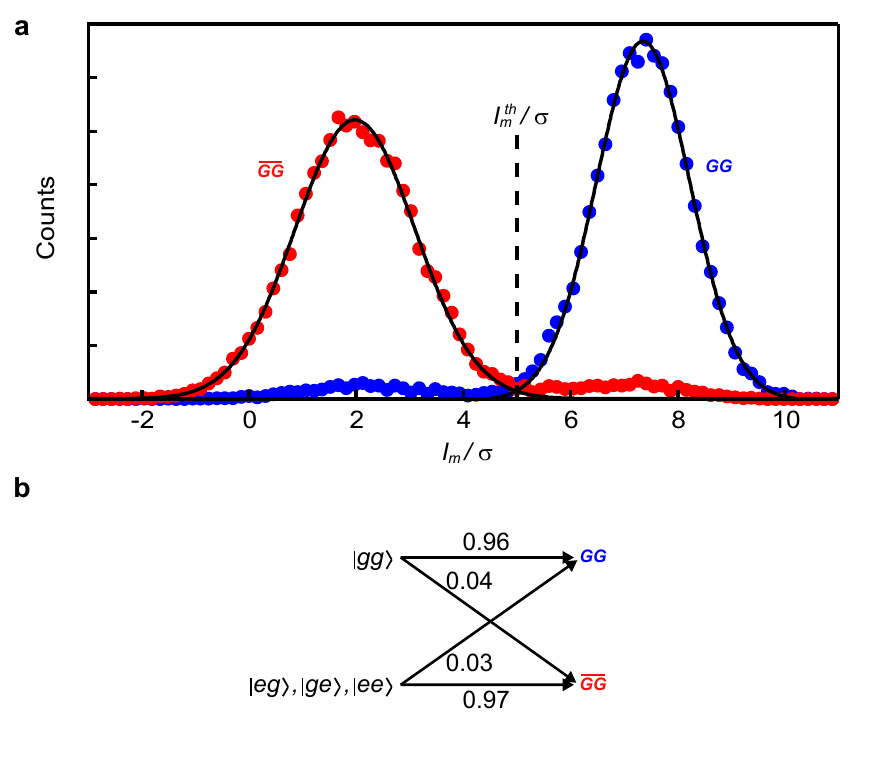}
\end{figure}
\textbf{Extended Data Figure 2: Single-shot readout of the observable $\proj{gg}$.}
\textbf{a.} Histogram of measurement outcomes recorded by the projective readout used for tomography. Outcome $I_m = 0$ implies that no microwave field was received in the $I$ quadrature for that measurement. Blue dots histogram labeled $GG$ was recorded with the qubits initially prepared in $\ket{gg}$ with a fidelity of $99.5$~\%. The red dots histogram labeled $\overline{GG}$ was recorded after identical preparation followed by a $\pi$-pulse on Alice. Solid lines are Gaussian fits. The horizontal axis of measurement outcomes $I_m$ is scaled by the average of the standard deviations of the two Gaussians, showing $5.5$ standard deviations between the centers of the two distributions. Dashed line indicates threshold that distinguishes $GG$ from $\overline{GG}$: an outcome $I_m > I_m^{th}$ is associated with $GG$ while $I_m < I_m^{th}$ is associated with $\overline{GG}$.
\textbf{b.} Summary of the fidelity of a single projective readout of the state of the two qubits assuming the separatrix $I_m^{th}/\sigma = 5$

\clearpage
\begin{figure}
\centering
\includegraphics{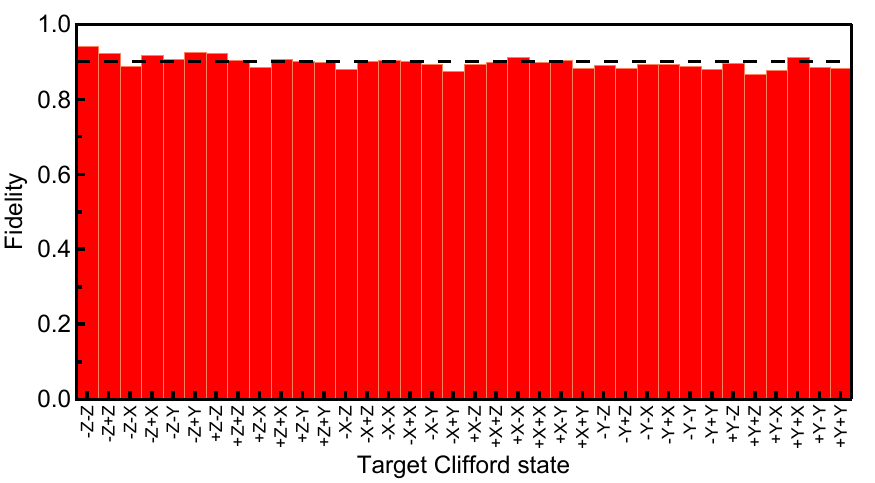}
\end{figure}
\textbf{Extended Data Figure 3: Calibration of systematic errors in tomography.}
Fidelity of two-qubit Clifford states measured by tomography identical to that used in the Bell state stabilization protocol. Clifford states are prepared by  starting in $\ket{gg}$ with fidelity of $99.5$~\% followed by individual single-qubit rotations.The fidelity varies from a maximum of $94$~\% for the state $\ket{-Z,-Z}$, to a minimum of $87$~\% for the state $\ket{+Y,+Z}$, averaging $90$~\% over the $36$ states (dashed line).

\clearpage
\begin{figure}
\centering
\includegraphics{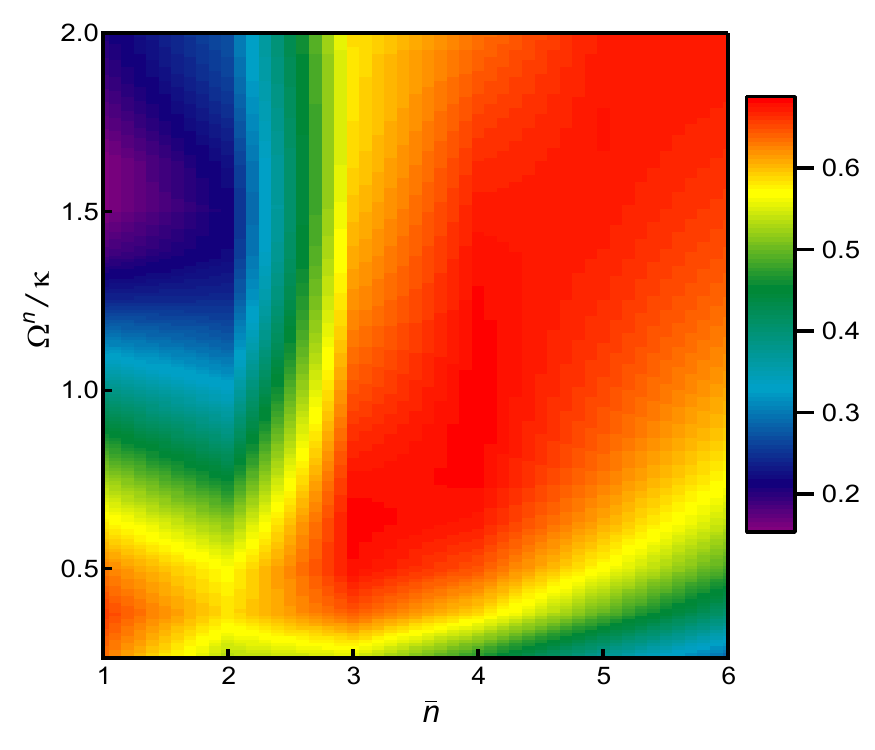}
\end{figure}
\textbf{Extended Data Figure 4: Predicted fidelity to $\ket{\phi_-}$ as a function of drive parameters $\bar n$ and $\Omega^n$ under the conditions of the present experiment.} $\Omega^0$ is taken to be $\kappa/2$ in this simulation. A broad distribution of parameter values resulting in a fidelity of about $70$~\%, indicates the robustness of the autonomous feedback protocol to variations in the drives.


\begin{thebibliography}{10}
\expandafter\ifx\csname url\endcsname\relax
  \def\url#1{\texttt{#1}}\fi
\expandafter\ifx\csname urlprefix\endcsname\relax\def\urlprefix{URL }\fi
\providecommand{\bibinfo}[2]{#2}
\providecommand{\eprint}[2][]{\url{#2}}

\bibitem{Nielsen2004}
\bibinfo{author}{Nielsen, M.~A.} \& \bibinfo{author}{Chuang, I.~L.}
\newblock \emph{\bibinfo{title}{Quantum Computation and Quantum Information}}
  (\bibinfo{publisher}{Cambridge University Press}, \bibinfo{year}{2004}).

\bibitem{Sayrin2011}
\bibinfo{author}{Sayrin, C.} \emph{et~al.}
\newblock \bibinfo{title}{Real-time quantum feedback prepares and stabilizes
  photon number states}.
\newblock \emph{\bibinfo{journal}{Nature}} \textbf{\bibinfo{volume}{477}},
  \bibinfo{pages}{73--77} (\bibinfo{year}{2011}).
\newblock \urlprefix\url{http://dx.doi.org/10.1038/nature10376}.

\bibitem{Vijay2012}
\bibinfo{author}{Vijay, R.} \emph{et~al.}
\newblock \bibinfo{title}{{Stabilizing Rabi oscillations in a superconducting
  qubit using quantum feedback}}.
\newblock \emph{\bibinfo{journal}{Nature}} \textbf{\bibinfo{volume}{490}},
  \bibinfo{pages}{77--80} (\bibinfo{year}{2012}).
\newblock \urlprefix\url{http://dx.doi.org/10.1038/nature11505}.

\bibitem{Riste2012b}
\bibinfo{author}{Rist\`e, D.}, \bibinfo{author}{Bultink, C.~C.},
  \bibinfo{author}{Lehnert, K.~W.} \& \bibinfo{author}{DiCarlo, L.}
\newblock \bibinfo{title}{Feedback control of a solid-state qubit using
  high-fidelity projective measurement}.
\newblock \emph{\bibinfo{journal}{Phys. Rev. Lett.}}
  \textbf{\bibinfo{volume}{109}}, \bibinfo{pages}{240502}
  (\bibinfo{year}{2012}).
\newblock
  \urlprefix\url{http://link.aps.org/doi/10.1103/PhysRevLett.109.240502}.

\bibitem{Campagne-Ibarcq2013}
\bibinfo{author}{Campagne-Ibarcq, P.} \emph{et~al.}
\newblock \bibinfo{title}{Persistent control of a superconducting qubit by
  stroboscopic measurement feedback}.
\newblock \emph{\bibinfo{journal}{Phys. Rev. X}} \textbf{\bibinfo{volume}{3}},
  \bibinfo{pages}{021008} (\bibinfo{year}{2013}).
\newblock \urlprefix\url{http://link.aps.org/doi/10.1103/PhysRevX.3.021008}.


\bibitem{Krauter2011}
\bibinfo{author}{Krauter, H.} \emph{et~al.}
\newblock \bibinfo{title}{Entanglement generated by dissipation and steady
  state entanglement of two macroscopic objects}.
\newblock \emph{\bibinfo{journal}{Phys. Rev. Lett.}}
  \textbf{\bibinfo{volume}{107}}, \bibinfo{pages}{080503}
  (\bibinfo{year}{2011}).
\newblock
  \urlprefix\url{http://link.aps.org/doi/10.1103/PhysRevLett.107.080503}.

\bibitem{Brakhane2012}
\bibinfo{author}{Brakhane, S.} \emph{et~al.}
\newblock \bibinfo{title}{Bayesian feedback control of a two-atom spin-state in
  an atom-cavity system}.
\newblock \emph{\bibinfo{journal}{Phys. Rev. Lett.}}
  \textbf{\bibinfo{volume}{109}}, \bibinfo{pages}{173601}
  (\bibinfo{year}{2012}).
\newblock
  \urlprefix\url{http://link.aps.org/doi/10.1103/PhysRevLett.109.173601}.

\bibitem{Riste2013}
\bibinfo{author}{{Rist{\`e}}, D.} \emph{et~al.}
\newblock \bibinfo{title}{{Deterministic entanglement of superconducting qubits by parity measurement and feedback}}.
\newblock \emph{\bibinfo{journal}{Nature}} \textbf{\bibinfo{volume}{502}},
  \bibinfo{pages}{350--354} (\bibinfo{year}{2013}).
\newblock \urlprefix\url{http://dx.doi.org/10.1038/nature12513}.

\bibitem{Geerlings2013}
\bibinfo{author}{Geerlings, K.} \emph{et~al.}
\newblock \bibinfo{title}{Demonstrating a driven reset protocol for a
  superconducting qubit}.
\newblock \emph{\bibinfo{journal}{Phys. Rev. Lett.}}
  \textbf{\bibinfo{volume}{110}}, \bibinfo{pages}{120501}
  (\bibinfo{year}{2013}).
\newblock
  \urlprefix\url{http://link.aps.org/doi/10.1103/PhysRevLett.110.120501}.

\bibitem{Murch2012}
\bibinfo{author}{Murch, K.~W.} \emph{et~al.}
\newblock \bibinfo{title}{Cavity-assisted quantum bath engineering}.
\newblock \emph{\bibinfo{journal}{Phys. Rev. Lett.}}
  \textbf{\bibinfo{volume}{109}}, \bibinfo{pages}{183602}
  (\bibinfo{year}{2012}).
\newblock
  \urlprefix\url{http://link.aps.org/doi/10.1103/PhysRevLett.109.183602}.

\bibitem{Barreiro2011}
\bibinfo{author}{Barreiro, J.~T.} \emph{et~al.}
\newblock \bibinfo{title}{An open-system quantum simulator with trapped ions}.
\newblock \emph{\bibinfo{journal}{Nature}} \textbf{\bibinfo{volume}{470}},
  \bibinfo{pages}{486--491} (\bibinfo{year}{2011}).
\newblock \urlprefix\url{http://dx.doi.org/10.1038/nature09801}.

\bibitem{Poyatos1996}
\bibinfo{author}{Poyatos, J.~F.}, \bibinfo{author}{Cirac, J.~I.} \&
  \bibinfo{author}{Zoller, P.}
\newblock \bibinfo{title}{Quantum reservoir engineering with laser cooled
  trapped ions}.
\newblock \emph{\bibinfo{journal}{Phys. Rev. Lett.}}
  \textbf{\bibinfo{volume}{77}}, \bibinfo{pages}{4728--4731}
  (\bibinfo{year}{1996}).
\newblock \urlprefix\url{http://link.aps.org/doi/10.1103/PhysRevLett.77.4728}.

\bibitem{Kerckhoff2010}
\bibinfo{author}{Kerckhoff, J.}, \bibinfo{author}{Nurdin, H.~I.},
  \bibinfo{author}{Pavlichin, D.~S.} \& \bibinfo{author}{Mabuchi, H.}
\newblock \bibinfo{title}{{Designing quantum memories with embedded control:
  Photonic circuits for autonomous quantum error correction}}.
\newblock \emph{\bibinfo{journal}{Phys. Rev. Lett.}}
  \textbf{\bibinfo{volume}{105}}, \bibinfo{pages}{040502}
  (\bibinfo{year}{2010}).
\newblock
  \urlprefix\url{http://link.aps.org/doi/10.1103/PhysRevLett.105.040502}.

\bibitem{Kastoryano2011}
\bibinfo{author}{Kastoryano, M.~J.}, \bibinfo{author}{Reiter, F.} \&
  \bibinfo{author}{S\o{}rensen, A.~S.}
\newblock \bibinfo{title}{Dissipative preparation of entanglement in optical
  cavities}.
\newblock \emph{\bibinfo{journal}{Phys. Rev. Lett.}}
  \textbf{\bibinfo{volume}{106}}, \bibinfo{pages}{090502}
  (\bibinfo{year}{2011}).
\newblock
  \urlprefix\url{http://link.aps.org/doi/10.1103/PhysRevLett.106.090502}.

\bibitem{Sarlette2011}
\bibinfo{author}{Sarlette, A.}, \bibinfo{author}{Raimond, J.~M.},
  \bibinfo{author}{Brune, M.} \& \bibinfo{author}{Rouchon, P.}
\newblock \bibinfo{title}{Stabilization of nonclassical states of the radiation
  field in a cavity by reservoir engineering}.
\newblock \emph{\bibinfo{journal}{Phys. Rev. Lett.}}
  \textbf{\bibinfo{volume}{107}}, \bibinfo{pages}{010402}
  (\bibinfo{year}{2011}).
\newblock
  \urlprefix\url{http://link.aps.org/doi/10.1103/PhysRevLett.107.010402}.

\bibitem{Lin2013}
\bibinfo{author}{Lin, Y.} \emph{et~al.}
\newblock \bibinfo{title}{Dissipative production of a maximally entangled
  steady state}.
\newblock \emph{\bibinfo{journal}{arXiv:1307.4443}}
  \urlprefix\url{http://arxiv.org/abs/1307.4443}.

\bibitem{Leghtas2013}
\bibinfo{author}{Leghtas, Z.} \emph{et~al.}
\newblock \bibinfo{title}{Stabilizing a bell state of two superconducting
  qubits by dissipation engineering}.
\newblock \emph{\bibinfo{journal}{Phys. Rev. A}} \textbf{\bibinfo{volume}{88}},
  \bibinfo{pages}{023849} (\bibinfo{year}{2013}).
\newblock \urlprefix\url{http://link.aps.org/doi/10.1103/PhysRevA.88.023849}.

\bibitem{Wallraff2004}
\bibinfo{author}{Wallraff, A.} \emph{et~al.}
\newblock \bibinfo{title}{Strong coupling of a single photon to a
  superconducting qubit using circuit quantum electrodynamics}.
\newblock \emph{\bibinfo{journal}{Nature}} \textbf{\bibinfo{volume}{431}},
  \bibinfo{pages}{162--167} (\bibinfo{year}{2004}).
\newblock \urlprefix\url{http://dx.doi.org/10.1038/nature02851}.

\bibitem{Schreier2008}
\bibinfo{author}{Schreier, J.~A.} \emph{et~al.}
\newblock \bibinfo{title}{Suppressing charge noise decoherence in
  superconducting charge qubits}.
\newblock \emph{\bibinfo{journal}{Phys. Rev. B}} \textbf{\bibinfo{volume}{77}},
  \bibinfo{pages}{180502} (\bibinfo{year}{2008}).
\newblock \urlprefix\url{http://link.aps.org/doi/10.1103/PhysRevB.77.180502}.

\bibitem{Nigg2012}
\bibinfo{author}{Nigg, S.~E.} \emph{et~al.}
\newblock \bibinfo{title}{Black-box superconducting circuit quantization}.
\newblock \emph{\bibinfo{journal}{Phys. Rev. Lett.}}
  \textbf{\bibinfo{volume}{108}}, \bibinfo{pages}{240502}
  (\bibinfo{year}{2012}).
\newblock
  \urlprefix\url{http://link.aps.org/doi/10.1103/PhysRevLett.108.240502}.

\bibitem{Schuster2007}
\bibinfo{author}{Schuster, D.~I.} \emph{et~al.}
\newblock \bibinfo{title}{Resolving photon number states in a superconducting
  circuit}.
\newblock \emph{\bibinfo{journal}{Nature}} \textbf{\bibinfo{volume}{445}},
  \bibinfo{pages}{515--518} (\bibinfo{year}{2007}).
\newblock \urlprefix\url{http://dx.doi.org/10.1038/nature05461}.

\bibitem{Lalumiere2010}
\bibinfo{author}{Lalumi\`ere, K.}, \bibinfo{author}{Gambetta, J.~M.} \&
  \bibinfo{author}{Blais, A.}
\newblock \bibinfo{title}{{Tunable joint measurements in the dispersive regime
  of cavity QED}}.
\newblock \emph{\bibinfo{journal}{Phys. Rev. A}} \textbf{\bibinfo{volume}{81}},
  \bibinfo{pages}{040301} (\bibinfo{year}{2010}).
\newblock \urlprefix\url{http://link.aps.org/doi/10.1103/PhysRevA.81.040301}.

\bibitem{Tornberg2010}
\bibinfo{author}{Tornberg, L.} \& \bibinfo{author}{Johansson, G.}
\newblock \bibinfo{title}{{High-fidelity feedback-assisted parity measurement
  in circuit QED}}.
\newblock \emph{\bibinfo{journal}{Phys. Rev. A}} \textbf{\bibinfo{volume}{82}},
  \bibinfo{pages}{012329} (\bibinfo{year}{2010}).
\newblock \urlprefix\url{http://link.aps.org/doi/10.1103/PhysRevA.82.012329}.

\bibitem{Paik2011}
\bibinfo{author}{Paik, H.} \emph{et~al.}
\newblock \bibinfo{title}{{Observation of high coherence in Josephson junction
  qubits measured in a three-dimensional circuit QED architecture}}.
\newblock \emph{\bibinfo{journal}{Phys. Rev. Lett.}}
  \textbf{\bibinfo{volume}{107}}, \bibinfo{pages}{240501}
  (\bibinfo{year}{2011}).
\newblock
  \urlprefix\url{http://link.aps.org/doi/10.1103/PhysRevLett.107.240501}.

\bibitem{Filipp2009}
\bibinfo{author}{Filipp, S.} \emph{et~al.}
\newblock \bibinfo{title}{Two-qubit state tomography using a joint dispersive
  readout}.
\newblock \emph{\bibinfo{journal}{Phys. Rev. Lett.}}
  \textbf{\bibinfo{volume}{102}}, \bibinfo{pages}{200402}
  (\bibinfo{year}{2009}).
\newblock
  \urlprefix\url{http://link.aps.org/doi/10.1103/PhysRevLett.102.200402}.

\bibitem{Bergeal2010a}
\bibinfo{author}{Bergeal, N.} \emph{et~al.}
\newblock \bibinfo{title}{{Phase-preserving amplification near the quantum
  limit with a Josephson ring modulator}}.
\newblock \emph{\bibinfo{journal}{Nature}} \textbf{\bibinfo{volume}{465}},
  \bibinfo{pages}{64--68} (\bibinfo{year}{2010}).
\newblock \urlprefix\url{http://dx.doi.org/10.1038/nature09035}.

\bibitem{Wootters1998}
\bibinfo{author}{Wootters, W.~K.}
\newblock \bibinfo{title}{Entanglement of formation of an arbitrary state of
  two qubits}.
\newblock \emph{\bibinfo{journal}{Phys. Rev. Lett.}}
  \textbf{\bibinfo{volume}{80}}, \bibinfo{pages}{2245--2248}
  (\bibinfo{year}{1998}).
\newblock \urlprefix\url{http://link.aps.org/doi/10.1103/PhysRevLett.80.2245}.

\bibitem{Slichter2012}
\bibinfo{author}{Slichter, D.~H.} \emph{et~al.}
\newblock \bibinfo{title}{{Measurement-induced qubit state mixing in circuit
  QED from up-converted dephasing noise}}.
\newblock \emph{\bibinfo{journal}{Phys. Rev. Lett.}}
  \textbf{\bibinfo{volume}{109}}, \bibinfo{pages}{153601}
  (\bibinfo{year}{2012}).
\newblock
  \urlprefix\url{http://link.aps.org/doi/10.1103/PhysRevLett.109.153601}.

\bibitem{Reichle2006}
\bibinfo{author}{Reichle, R.} \emph{et~al.}
\newblock \bibinfo{title}{Experimental purification of two-atom entanglement}.
\newblock \emph{\bibinfo{journal}{Nature}} \textbf{\bibinfo{volume}{443}},
  \bibinfo{pages}{838--841} (\bibinfo{year}{2006}).
\newblock \urlprefix\url{http://dx.doi.org/10.1038/nature05146}.

\bibitem{Leghtas2012}
\bibinfo{author}{Leghtas, Z.} \emph{et~al.}
\newblock \bibinfo{title}{Hardware-efficient autonomous quantum memory protection}.
\newblock \emph{\bibinfo{journal}{Phys. Rev. Lett.}}
  \textbf{\bibinfo{volume}{111}}, \bibinfo{pages}{120501}
  (\bibinfo{year}{2013}).
\newblock
  \urlprefix\url{http://link.aps.org/doi/10.1103/PhysRevLett.111.120501}.

\setcounter{firstbib}{\value{enumiv}}

\end{thebibliography}

\begin{thebibliography}{10}
\expandafter\ifx\csname url\endcsname\relax
  \def\url#1{\texttt{#1}}\fi
\expandafter\ifx\csname urlprefix\endcsname\relax\def\urlprefix{URL }\fi
\providecommand{\bibinfo}[2]{#2}
\providecommand{\eprint}[2][]{\url{#2}}

\setcounter{enumiv}{\value{firstbib}}

\bibitem{Lecocq2011}
\bibinfo{author}{Lecocq, F.} \emph{et~al.}
\newblock \bibinfo{title}{Junction fabrication by shadow evaporation without a
  suspended bridge}.
\newblock \emph{\bibinfo{journal}{Nanotechnology}}
  \textbf{\bibinfo{volume}{22}}, \bibinfo{pages}{315302}
  (\bibinfo{year}{2011}).
\newblock \urlprefix\url{http://stacks.iop.org/0957-4484/22/i=31/a=315302}.

\bibitem{Rigetti2009}
\bibinfo{author}{Rigetti, C.}
\newblock \emph{\bibinfo{title}{Quantum Gates for Superconducting Qubits}}.
\newblock Ph.D. thesis, \bibinfo{school}{Yale University},
  \bibinfo{address}{New Haven, Connecticut, USA} (\bibinfo{year}{2009}).

\bibitem{Houck2008}
\bibinfo{author}{Houck, A.~A.} \emph{et~al.}
\newblock \bibinfo{title}{Controlling the spontaneous emission of a
  superconducting transmon qubit}.
\newblock \emph{\bibinfo{journal}{Phys. Rev. Lett.}}
  \textbf{\bibinfo{volume}{101}}, \bibinfo{pages}{080502}
  (\bibinfo{year}{2008}).
\newblock
  \urlprefix\url{http://link.aps.org/doi/10.1103/PhysRevLett.101.080502}.

\bibitem{Sears2012}
\bibinfo{author}{Sears, A.~P.} \emph{et~al.}
\newblock \bibinfo{title}{{Photon shot noise dephasing in the strong-dispersive
  limit of circuit QED}}.
\newblock \emph{\bibinfo{journal}{Phys. Rev. B}} \textbf{\bibinfo{volume}{86}},
  \bibinfo{pages}{180504} (\bibinfo{year}{2012}).
\newblock \urlprefix\url{http://link.aps.org/doi/10.1103/PhysRevB.86.180504}.

\bibitem{Hatridge2013}
\bibinfo{author}{Hatridge, M.} \emph{et~al.}
\newblock \bibinfo{title}{Quantum back-action of an individual
  variable-strength measurement}.
\newblock \emph{\bibinfo{journal}{Science}} \textbf{\bibinfo{volume}{339}},
  \bibinfo{pages}{178} (\bibinfo{year}{2013}).
\newblock \urlprefix\url{http://dx.doi.org/10.1126/science.1226897}.

\bibitem{Johnson2012}
\bibinfo{author}{Johnson, J.~E.} \emph{et~al.}
\newblock \bibinfo{title}{Heralded state preparation in a superconducting
  qubit}.
\newblock \emph{\bibinfo{journal}{Phys. Rev. Lett.}}
  \textbf{\bibinfo{volume}{109}}, \bibinfo{pages}{050506}
  (\bibinfo{year}{2012}).
\newblock
  \urlprefix\url{http://link.aps.org/doi/10.1103/PhysRevLett.109.050506}.

\bibitem{Chow2010}
\bibinfo{author}{Chow, J.~M.} \emph{et~al.}
\newblock \bibinfo{title}{Detecting highly entangled states with a joint qubit
  readout}.
\newblock \emph{\bibinfo{journal}{Phys. Rev. A}} \textbf{\bibinfo{volume}{81}},
  \bibinfo{pages}{062325} (\bibinfo{year}{2010}).
\newblock \urlprefix\url{http://link.aps.org/doi/10.1103/PhysRevA.81.062325}.

\bibitem{Motzoi2009}
\bibinfo{author}{Motzoi, F.}, \bibinfo{author}{Gambetta, J.~M.},
  \bibinfo{author}{Rebentrost, P.} \& \bibinfo{author}{Wilhelm, F.~K.}
\newblock \bibinfo{title}{Simple pulses for elimination of leakage in weakly
  nonlinear qubits}.
\newblock \emph{\bibinfo{journal}{Phys. Rev. Lett.}}
  \textbf{\bibinfo{volume}{103}}, \bibinfo{pages}{110501}
  (\bibinfo{year}{2009}).
\newblock
  \urlprefix\url{http://link.aps.org/doi/10.1103/PhysRevLett.103.110501}.

\bibitem{Reed2013}
\bibinfo{author}{Reed, M.~D.}
\newblock \emph{\bibinfo{title}{Entanglement and Quantum Error Correction with
  Superconducting Qubits}}.
\newblock Ph.D. thesis, \bibinfo{school}{Yale University},
  \bibinfo{address}{New Haven, Connecticut, USA.} (\bibinfo{year}{2013}).

\bibitem{Devoret2013}
\bibinfo{author}{Devoret, M.~H.} \& \bibinfo{author}{Schoelkopf, R.~J.}
\newblock \bibinfo{title}{{Superconducting circuits for quantum information: An
  outlook}}.
\newblock \emph{\bibinfo{journal}{Science}} \textbf{\bibinfo{volume}{339}},
  \bibinfo{pages}{1169--1174} (\bibinfo{year}{2013}).
\newblock \eprint{http://www.sciencemag.org/content/339/6124/1169.full.pdf}.

\bibitem{Itano1990}
\bibinfo{author}{Itano, W. M.}, \bibinfo{author}{Heinzen, D. J.},
  \bibinfo{author}{Bollinger, J. J.} \& \bibinfo{author}{Wineland, D. J.}
\newblock \bibinfo{title}{Quantum Zeno effect}.
\newblock \emph{\bibinfo{journal}{Phys. Rev. A}}
  \textbf{\bibinfo{volume}{41}}, \bibinfo{pages}{2295}
  (\bibinfo{year}{1990}).
\newblock \urlprefix\url{http://link.aps.org/doi/10.1103/PhysRevA.41.2295}.

\bibitem{Reed2010}
\bibinfo{author}{Reed, M.~D.} \emph{et~al.}
\newblock \bibinfo{title}{Fast reset and suppressing spontaneous emission of a
  superconducting qubit}.
\newblock \emph{\bibinfo{journal}{Appl. Phys. Lett.}}
  \textbf{\bibinfo{volume}{96}}, \bibinfo{pages}{203110}
  (\bibinfo{year}{2010}).
\newblock \urlprefix\url{http://link.aip.org/link/?APL/96/203110/1}.

\end{thebibliography}
\end{document}